\documentclass[10pt]{article}
\pdfoutput=1
\usepackage[utf8]{inputenc}
\usepackage{amssymb}
\usepackage{graphicx}
\usepackage{graphics}
\usepackage{amsmath}
\def\const{\mbox{const}}
\def\e{{\rm e}}

\def\g{\gamma}

\def\l{\left(}
\def\r{\right)}

\newcommand{\be}{\begin{equation}}
\newcommand{\ee}{\end{equation}}
\newcommand{\ba}{\begin{align}}
\newcommand{\ea}{\end{align}}
\newcommand{\bg}{\begin{gather}}
\newcommand{\eg}{\end{gather}}
\newcommand{\bseq}{\begin{subequations}}
\newcommand{\eseq}{\end{subequations}}

\makeatletter
\def\gsim{\compoundrel>\over\sim}
\def\lsim{\compoundrel<\over\sim}
\def\compoundrel#1\over#2{\mathpalette\compoundreL{{#1}\over{#2}}}
\def\compoundreL#1#2{\compoundREL#1#2}
\def\compoundREL#1#2\over#3{\mathrel
         {\vcenter{\hbox{$\m@th\buildrel{#1#2}\over{#1#3}$}}}}
\makeatother

\begin{document}
\title{Light hidden photon production
   in high energy collisions
}
\author{S. Demidov${}^{1,2}$, S. Gninenko${}^{1}$ and D. Gorbunov${}^{1,2}$\\
${}^1$ 
  Institute for Nuclear Research of the Russian Academy of Sciences, 
Moscow 117312, Russia\\
${}^2$ Moscow Institute of Physics and Technology, 
Dolgoprudny 141700, Russia\\ 
  E-mails:  \texttt{demidov@ms2.inr.ac.ru},
  \texttt{Sergei.Gninenko@cern.ch}, \texttt{gorby@ms2.inr.ac.ru}
}

\date{}

\maketitle

\begin{abstract} 
The visible and dark sectors of particle physics can be connected via 
the kinetic mixing between the ordinary ($\g$) and hidden photon
($\g'$). If the latter is 
light its production in high energy collisions of ordinary particles 
occurs 
 via the $\g - \g'$ oscillations 
similarly to the neutrino
processes. Generically, the experiments are insensitive to mass of the
hidden photon, if it is lighter than 1\,MeV, and it does not decay into 
$e^+e^-$-pair. Still, one can use the missing energy and scattering
off the detector as signatures to search for light  hidden photon. 
Presence of media suppresses 
production of the light vectors making the experiments insensitive to
the entire model. 
In media the light hidden photon production  is  typically suppressed
due to the dump of the $\g - \g'$  oscillations  making the experiments insensitive to
the entire model. 
We present analytic formulas for the light hidden
photon production, propagation and detection valid for searches at colliders and
beam-target experiments and apply them to estimate the impact on the
sensitivities of NA64, FASER, MATHUSLA, SHiP, T2K, DUNE and
  NA62 for the background-free case. 
\end{abstract}

\section{Introduction}

New physics can involve GeV-scale and even lighter particles, the
pattern argued to be preferable when the gauge hierarchy problem is
addressed \cite{Vissani:1997ys,Giudice:2013nak}. 
These new particles can be directly 
searched for in high-intensity collision experiments operating at
collider or fixed-target facilities. They include universal
experiments like those running at the LHC, long-baseline neutrinos oscillation
experiments exploiting proton beams incident on target (DUNE, T2K) and
specially designed projects with high-intensity beams dedicated to
new physics hunting (NA64, SHiP, MATHUSLA). Generically, whenever very
light particles are concerned, astrophysical processes exhibit a higher
sensitivity than the direct searches, but the latter naturally earn a higher
confidence, and we concentrate on them.

The new particles, if light, must be singlet with respect to the gauge
group of the Standard Model of particle physics (SM). Otherwise,
absence of their traces in numerous experiments performed so far
would be difficult to explain. If the new singlet is a Lorenzian vector, 
$X_\mu$, its most natural effective interaction with the SM fields is via
so-called vector portal --- renormalizable mixing between the
gauge-invariant kinetic
terms of $X_\mu$ and photon $A_\mu$. Below the electroweak scale around
100\,GeV the corresponding lagrangian reads
\be
\label{main-lagr}
{\cal L} = -\frac{1}{4}F_{\mu\nu}F^{\mu\nu} -
\frac{1}{4}X_{\mu\nu}X^{\mu\nu} -
\frac{\epsilon}{2}F_{\mu\nu}X^{\mu\nu}
+ \frac{1}{2}m_X^2X_\mu X^\mu + A_\mu j^\mu\;,
\ee
where $m_X$ is the hidden photon mass emerged presumably due to the
Higgs mechanism operating in the hidden sector. The kinetic mixing
with the ordinary photon in \eqref{main-lagr} solely defines the hidden photon
phenomenology associated with the SM particles: $X_\mu$ can be
produced by a virtual photon and can decay into the SM particles via
a virtual photon wherever kinematically allowed. The hidden photon can
also effectively scatter off the electrically charged SM particles
because of this mixing. Besides, in the hidden sector $X_\mu$ can couple to 
dark matter particles \cite{Pospelov:2007mp} hence forming a
gate between the visible and dark Worlds. From phenomenological side,
the couplings to particles from the hidden sector imply a
possibility of invisible decay mode for $X_\mu$, which alters the
strategy of its searches \cite{Ilten:2018crw}.

In this paper we study phenomenology of the hidden photon,
lighter than 1\,MeV, when the decay mode into $e^+e^-$ is forbidden.  And
we take the limit where the hidden photons participate in processes
(e.g. emerge, scatter, etc) being ultra-relativistic in the laboratory
frame, $E\gg m_X$. Then in many cases
the interaction of the hidden photon with the SM particles happens via
oscillations $X_\mu\leftrightarrow A_\mu$, rather than via the Compton
scattering (see e.g.~\cite{Ahlers:2007rd}). In particular, applying oscillations to describe the
hidden photon behavior in a nuclear reactor it has been recently
found\,\cite{Danilov:2018bks} that, generically, there are no chances
to produce too light hidden photon there. The reason is that the
visible photon either gets absorbed or scatters in media before
oscillating into the hidden photon. The electromagnetic interactions
of the visible photon in media effectively suppress the oscillations
with respect to the vacuum case. The only exception is a resonance
region, where the hidden photon mass coincides with the plasma
frequency in the
media, where the photons propagate.

In the main part of the paper we give formulas to describe the 
visible photon conversion to the hidden photon and back which could be useful for several classes of 
experiments. 
Usually the accelerator experiments do not provide bounds on the
visible-to-hidden photon mixing for $m_X<1$\,MeV, unless specially
designed. However, in many cases they can extend the limits to this
region, and our formulas will help to do it.  
We discuss their applicability and prospects of
these experiments to explore the models with the light hidden photon. 

The plan of the paper is as follows.
We present the detailed description of the oscillations in
Sec.\,\ref{sec:oscillations}. In Sec.\,\ref{sec:production} we apply
this approach  to calculate the light hidden photon 
production probability for the cases of collisions in vacuum (e.g. a collider) and in matter
(e.g. a beam-dump experiment). Generally, the hidden photon can be
detected either via disappearance of the visible photon or via 
appearance of a visible photon from nowhere (``light shining through the
wall''). The both processes for the light ultrarelativistic hidden photon
are described as hidden-to-visible photon oscillations and 
discussed in detail in Sec.\,\ref{sec:registration}. Finally, in
Sec.\,\ref{sec:examples} we estimate the conversion probability 
for several presently ongoing and future experiments,
and investigate the impact of our findings on their prospects in probing models with light hidden
photon assuming zero background and 100\% detection efficiency (that implies the most optimistic
numbers). We summarize in Sec.\,\ref{sec:summary} emphasizing the
necessary conditions for the resonance amplification of the
hidden photon production in accelerator experiments.

\section{Oscillations between visible and hidden photons}
\label{sec:oscillations}

In this Section we present main formulas describing oscillations of
the visible--hidden photon system in vacuum and media.

In the interesting case of small mixing, $\epsilon\ll1$, it is
convenient to replace $X_\mu$ with $S_\mu\equiv X_\mu+\epsilon A_\mu$,
so that in the new variables the kinetic term in lagrangian
\eqref{main-lagr} is diagonal, but mixing emerges in the mass term,
instead. State $S_\mu$ remains sterile with respect to the SM gauge
interactions and we call it the hidden photon state. The system evolution
written in terms of the new fields allows for a very simple description
via oscillations. Namely, while the electromagnetic current
$j_\mu$ produces a quantum wave-packet of photon $A_\mu$, the states
which propagate --- eigenvectors of the Hamiltonian --- are mixtures
of $A_\mu$ and $S_\mu$. As it follows from~\eqref{main-lagr} the
Hamiltonian describing the system ($A,S$) 
in the ultrarelativistic regime in vacuum has the form
\be
\label{ham-prop}
H = \frac{1}{2E}
\left(
\begin{array}{cc}
      0 & -\epsilon m_X^2\\
  -\epsilon m_X^2 & m_X^2
\end{array}
\right)
\ee
up to corrections of order ${\mathbb O}(\epsilon^2)$
in its diagonal part. Here $E$ is the photon energy.
Note in passing that we consider here only two transverse polarizations of 
the massive hidden photon. As for the third, longitudinal polarization, it
is out of interest here, because its production in the processes under
discussion is suppressed, as compared to the transverse polarization
modes, by factor $m_X^2/E^2\ll 1$, see 
Ref.\,\cite{Redondo:2013lna} for the details.  
The corresponding time evolution between the ``interaction eigenstates'',
i.e. $A$ and $S$, is very similar to the oscillations between neutrino
flavours.  To the leading order in $\epsilon$ one obtains from \eqref{ham-prop} for the
transition probability from visible to hidden photon at a distance $L$
from the source,  
\begin{equation}
  \label{vacuum-probability}
P\l\gamma\to\gamma'\r= 4\,\epsilon^2\sin^2\l \frac{\delta m^2 L}{4\,E} \r\,
\end{equation}
with $\delta m^2 \equiv m_X^2$, which is a replica of the
neutrino vacuum oscillation probability. From
Eq.~\eqref{vacuum-probability} one can define the oscillation length as   
\begin{equation}
  \label{osc-length}
  L_{osc}=\frac{4\pi E}{\delta m^2}\approx
  2.5\,\text{cm}\,\frac{E}{1\,\text{MeV}}\frac{\l 10\,\text{eV}\r^2}{\delta m^2}\;,
\end{equation}
which is an important characteristic of this process.

Similarly to the neutrino case for the visible and hidden photons to oscillate,  several
coherence conditions must be fulfilled, see
e.g.~\cite{Kayser:1981ye}. Firstly, size of the initial 
wave packet $\sigma_x$ of the produced photon should be smaller than
the oscillation length, i.e. $\sigma_x<L_{osc}$. At the same time
the oscillations terminate when individual wave packets of the mass
eigenstates become spatially separated due to the difference in their
velocities $\Delta v$. This happens at a coherence distance $l_{coh}$,
which can 
be estimated as 
\begin{equation}
  \label{l_coh}
  l_{coh}\approx \frac{\sigma_x}{\Delta v} = \frac{2\sigma_x
    E^2}{\delta m^2}\;.
\end{equation}
The size of the photon wave packet strongly depends on the visible photon production
process. For instance, for the case of direct photon production in
the Compton scattering one expects
\be
\sigma_x\sim \frac{1}{q},
\ee
where $q$ is the transfer momentum in the reaction. For the photons produced
in decays of neutral pions $\pi^0$ propagating in vacuum with energy $E$ one can
obtain the following estimate  
\be
\label{pi0-wwidth}
\sigma_x\sim \frac{E}{m_{\pi^0}}\tau_{\pi^0} \sim 2.6\times
10^{-6}\times \frac{E}{m_{\pi^0}}\,{\rm cm}\;,
\ee
where $\tau_{\pi^0}$ is the lifetime of neutral pion in its rest
frame.

Once the coherence gets lost and in particular if the photon source is not
monochromatic or is not compact as compared to the oscillation length,
one can average the oscillating factor in \eqref{vacuum-probability}
and arrive at the simple expression 
\be
\label{prob-av-vacuum}
P\l\gamma\to\gamma'\r_{av}= 2\,\epsilon^2\,.
\ee

When the system wave packet propagates in media, the visible photon interacts with the
environment. This results in several important consequences which
should be taken into account when considering time evolution of the
system in question. First of all, the photon forward scattering off (free)
electrons 
in the media results in a modification of its dispersion relation, see
e.g.~\cite{ll10,Braaten:1993jw,Raffelt:1996wa}. The corresponding change for the transverse polarizations looks as if 
these modes get an effective mass. For the processes under
discussion the value of this mass coincides with the plasma frequency,
i.e. 
\begin{equation}
  \label{photon-mass}
m_\gamma^2=4\pi\alpha\frac{n_e}{m_e}\;,
\end{equation}
where $\alpha$ is the fine structure constant and $n_e$ is the 
density of free electrons along the photon trajectory. 
Second, the visible photon can be absorbed and/or rescattered. This
results in an additional source of coherence 
loss because, if a state, which is a mixture of $A_\mu$ and $S_\mu$,
endures either absorption or (in)elastic scattering; the latter brings
the state back to the pure visible photon state which  
starts its time evolution. The corresponding
interaction length
$1/\Gamma$ is determined by the material of the media.
We somewhat loosely call $1/\Gamma$ as attenuation, interaction and
absorption length through the paper, having in mind its meaning: it 1)
terminates the oscillations and 2) can be adopted as a signature of
converted back visible photon in the detector. 
Full
description of the system evolution can be obtained applying the 
density matrix formalism. In what follows we do not take into 
account rescattered photons\footnote{We consider them as being absorbed at initial energy
and reappeared at a smaller energy.} in a single formula (all of them must be
sum up separately) and thus both effects --- 
the effective mass and interactions with the media --- can be taken into
account by the following modification of  $H_{11}$ component of the Hamiltonian
matrix~\eqref{ham-prop}: 
\be
\frac{\epsilon m_X^2}{2E}\to \frac{\epsilon m_X^2 + m_\gamma^2}{2E} -
i\frac{\Gamma}{2}\;.
\ee
It is convenient to rewrite the Hamiltonian in the form
\be
\label{ham}
H = \left(
\begin{array}{cc}
  -i\gamma  & \delta \\
  \delta & {\cal E}
\end{array}\right)
\ee
by subtracting a part proportional to the unity matrix.
Here $\delta  \equiv \epsilon m_X^2/(2 E)$ is the mixing parameter, ${\cal 
  E}\equiv\Delta m^2/(2E)$, with $\Delta m^2 =
m_X^2 - m_\gamma^2$, and $\gamma\equiv \Gamma/2$ is
the parameter describing attenuation of the photon flux $F_\gamma$ at a given
energy due to
interaction with matter. This Hamiltonian can be used to determine
the time evolution of the system. Note in passing, that the Hamiltonian
  of the form~\eqref{ham} is common for studying of two-level systems like e.g.
  neutron-antineutron oscillations~\cite{Kerbikov:2018mct}.

For the simplest case of homogeneous
media the solution to the corresponding Schrodinger equation can be found
explicitly and is presented in Appendix\,\ref{App:Solution}. It has relatively simple form in two limiting cases: 1)
when $\epsilon$ is sufficiently small, namely if
$\left|\frac{\delta}{{\cal E}+i\gamma}\right|\ll 1$; 2) when the
mixing is almost maximal, i.e. $\left|\frac{\delta}{{\cal
    E}+i\gamma}\right|\gg 1$.
In the first case, assuming additionally $\frac{\delta^2}{{\cal
    E}^2+\gamma^2}\gamma L\ll 1$ which is fulfilled in the examples below
 (where $L$ is the distance covered by the propagating state),  
we obtain for the pure photon initial state
\be
\psi(0) \equiv \left(
\begin{array}{c}
  1 \\
  0
\end{array}\right) \to
\psi(L)\equiv \left(
\begin{array}{c}
  \psi_1(L) \\
  \psi_2(L)
\end{array}
\right)
\approx \left(
\begin{array}{c}
  {\rm e}^{-\gamma L} 
  \\
  - \frac{\delta}{{\cal E} +
    i\gamma}\left({\rm e}^{-\gamma L} - {\rm e}^{-i{\cal E}L}\right)
\end{array}
\right)
\label{matu}
\ee
and the transition probability $P\l\gamma\to\gamma'\r=\left|\psi_2(L)
\right|^2$ reads\,\cite{Redondo:2015iea} 
\begin{equation}
  \label{probability-homogeneous}
  P\l\gamma\to\gamma'\r=
  \frac{\delta^2}{{\cal E}^2+\gamma^2}\l 1 +  \e^{-2 \gamma L}-2\,\e^{-\gamma
    L}\cos\l {\cal E} L\r\r  = 
  \frac{\epsilon^2m^4}{\l\Delta
  m^2\r^2+E^2\Gamma^2}\l 1+ \e^{-\Gamma L}-2\,\e^{-\frac{\Gamma
    L}{2}}\cos\l \frac{\Delta m^2 L}{2\,E} \r\r\,.
\end{equation}
The corresponding oscillation length can be found from~\eqref{osc-length}
with $\delta m^2\equiv \Delta m^2$.

In the case of inhomogeneous media the transition
amplitude between visible and hidden photons to the
leading order in $\delta$ 
is~\cite{Redondo:2015iea,Raffelt:1987im}  
\begin{equation}
  \label{general-amplitude}
  A\l \gamma\to\gamma'\r = \delta\int_0^L dl
  \e^{-i\int_0^l dl' {\cal E}(l')-\int_0^l dl'
    \gamma(l')}\,,
\end{equation}
where all the integrals are taken along the photon trajectory. The
transition probability is then obtained as 
\be
\label{general-probability}
P\l\gamma\to\gamma'\r = \left|A\l \gamma\to\gamma'\r\right|^2\;.
\ee

For propagation in the media further simplification in the description
happens whenever the distance to the source considerably exceeds the
absorption length, i.e. $L\gg 1/\Gamma$. Then the transition probability
\eqref{probability-homogeneous} reduces to distance-independent
formula
\begin{equation}
  \label{probability-constant}
P\l\gamma\to\gamma'\r= \frac{\epsilon^2m^4}{\l\Delta
  m^2\r^2+E^2\Gamma^2}\,.
\end{equation}
At such distances oscillations stop simply because of the photon
absorption, and probability to observe the hidden photon
\eqref{probability-homogeneous}  approaches
the constant value given by \eqref{probability-constant}. Let us note that in the case when
  $\frac{\delta^2}{{\cal E}^2 + \gamma^2}\gamma L\gsim 1$ the probability~\eqref{probability-constant}
is multiplied by suppression factor ${\rm e}^{-2\frac{\delta^2}{{\cal E}^2+\gamma^2}\gamma L}$.

In the case of almost maximal mixing, i.e. when
$\left|\frac{\delta}{{\cal E}+i\gamma}\right|\gg 1$, the
visible-to-hidden photon oscillation probability reads
\be
\label{prob-res}
P\l\gamma\to\gamma'\r={\rm e}^{-\gamma L}\sin^2\l\delta L\r.
\ee
This expression does not contain the small factor $\epsilon^2$
which suppresses the probability of hidden photon production in
Eqs.~\eqref{probability-homogeneous} and~\eqref{probability-constant}.

The amplitude of the inverse process --- conversion
from hidden to visible photon --- can be obtained in a similar
fashion. This process is important for the detection of a signal from
the hidden photons. Let us note that the detection occurs precisely
due to photon interaction in the media and we  can consider the attenuation
of the photon flux in the detector as a signature of the hidden
photon.

Typically, to detect a hidden photon signal one bears in mind either
missing energy signature (like e.g. in NA64 experiment) or
``shining-through-the-wall'' type of experiments, i.e. a setup where
production and detection regions are separated by a media in which all
visible photons would be absorbed. In the former case the typical
observable, e.g. number of disappeared photons, scales as
$\epsilon^2$. In the latter case, generally, the probability to
observe visible photon in the detector after it has been initially
produced in a source can be found by solving corresponding Schrodinger
equation along the photon trajectory. However, in most of the
experimental setups to be discussed below can often describe the
process in question as one consisting of two stages. During the first
stage a flux of hidden photons is produced outside the production
region. At the second stage the hidden photons are converted into
visible ones in the detector. The signal probability can be
approximated by a product of the probabilities of visible-to-hidden
and hidden-to-visible photon conversions. According to the above
discussion the signal is expected to scale as the fourth power of the
mixing parameter $\epsilon$. We investigate the both probabilities in
details in the next two Sections.

\section{Production of hidden photons}
\label{sec:production}

To calculate the light hidden photon production in a given experiment
one must sum up all the photons emerging from various sources: direct 
bremsstrauhlung, hadron decays, nuclear deexcitation, etc.
The key difference with the neutrino oscillations is the fact that the hidden
photon state $S_\mu$ is sterile with respect to direct interactions
with the SM particles. 

Wherever the photons appear, they can be converted into hidden photons
(as described in the previous Section) with the probability largely
depending on the photon mass, energy and environment. Besides,
interaction of 
the photon with the media results in production of the secondary photons
(i.e. in electromagnetic showers) which in turn can be converted into 
the hidden photons during their evolution. As we mentioned in
Sec.\,\ref{sec:oscillations}, to obtain the full description one
should use the density
matrix. This goes beyond the scope of this paper, where we just
outline the oscillation probabilities for the relevant mass ranges where
the oscillation description should be valid.

Produced visible photons start propagating in vacuum (low density
region) or in media (high density region). 
The vacuum case is naturally realized at colliders, when the region
close to the collision point is (almost) empty, material free. Then,
depending on the beam type, one can sum up the bremshtrauhlung
contribution and meson decays into photons, and convolute it with the
oscillation probability \eqref{vacuum-probability} taking into account
the coherence conditions.
Eventually, photon states reach regions with a dense material
(e.g. detectors), and the oscillation dynamics changes
accordingly. Hence, if the produced photon 
covers the distance $d$, the system wave function evolves as \eqref{matu} 
(set $L=d$ and $\gamma=0$),
which gives \eqref{vacuum-probability} for the oscillation
probability $P=|\psi_2(d)|^2$.

In the general case of propagation in matter, to find the oscillation
probability one must use \eqref{general-probability}, numerically
integrating along and summing over all the photon trajectories.
However, in many cases the media can be described as a set of layers of different
but homogeneous media. Then the hidden photon production can be estimated
analytically by solving the Schrodinger equation inside each layer
with help of \eqref{exact} and subsequent matching of the results at layer
borders.  In particular, in the first case, to the leading nontrivial
order in $\delta$ (or equivalently $\epsilon$) one finds the wave
function at a distance $L$
\be
\psi(0) \equiv \left(
\begin{array}{c}
  a \\
  b
\end{array}\right) \to
\psi(L) \approx \left(
\begin{array}{c}
  a{\rm e}^{-\gamma L} - \frac{\delta}{{\cal E} + i\gamma}b\left({\rm
    e}^{-\gamma L} - {\rm e}^{-i{\cal E}L}\right)
  \\
  b{\rm e}^{-i{\cal E}L} - \frac{\delta}{{\cal E} +
    i\gamma}a\left({\rm e}^{-\gamma L} - {\rm e}^{-i{\cal E}L}\right)
\end{array} 
\right)
\equiv U(L)\psi(0)\;.
\label{uU}
\ee
In the last
expression we introduce  the evolution operator $U(L)$ which is used
in what follows.

After the photon passing through the first layer (with
corresponding parameters $\gamma_1$, ${\cal E}_1$ and layer width
$d_1$) one finds from \eqref{uU}
the evolution operator 
  \begin{equation}
    \label{propagate}
U_1(d_1)=\begin{pmatrix}
\e^{-\gamma_1 d_1}  & -\frac{\delta}{{\cal E}_1+i\gamma_1}\l{\rm
  e}^{-\gamma_1 d_1} - {\rm e}^{-i{\cal E}_1d_1}\r\\
-\frac{\delta}{{\cal E}_1+i\gamma_1}
\l \e^{-\gamma_1 d_1} - \e^{-i{\cal E}_1 d_1}\r & \e^{-i {\cal E}_1 d_1} 
\end{pmatrix}.
\end{equation}
After passing the second layer one gets 
\be
\label{vac-red}
\psi(d)= U_2(d_2)U_1(d_1) \left(
\begin{array}{c}
  1 \\
  0
\end{array}\right)
= \left(
\begin{array}{c}
  {\rm e}^{-\gamma_1d_1-\gamma_2d_2}
  \\
  - \frac{\delta}{{\cal
      E}_1+i\gamma_1}
  \l {\rm e}^{-\gamma_1d_1} - {\rm e}^{-i{\cal
      E}_1d_1}\r {\rm e}^{-i{\cal E}_2d_2} - \frac{\delta}{{\cal
    E}_2+i\gamma_2} \l \text{e}^{-\gamma_2 d_2} - {\rm e}^{-i{\cal
      E}_2d_2}\r {\rm e}^{-\gamma_1d_1} 
\end{array}
\right),
\ee
where $d=d_1+d_2$ and we neglected contributions of order $\delta^2$. 
In particular, this result describes the evolution in the case of collider
setup, when the system starts from photon propagating in vacuum and then enters
the media. Taking the first layer as vacuum ($\gamma_1=0$) one finds from
\eqref{vac-red}
\be
\label{vac-mat}
\psi(d)= U_2(d_2)U_1(d_1) \left(
\begin{array}{c}
  1 \\
  0
\end{array}\right)
= \left(
\begin{array}{c}
  {\rm e}^{-\gamma_2d_2}
  \\
  - \frac{\delta}{{\cal
      E}_1}
  \l 1 - {\rm e}^{-i{\cal
      E}_1d_1}\r {\rm e}^{-i{\cal E}_2d_2} - \frac{\delta}{{\cal
    E}_2+i\gamma_2} \l \text{e}^{-\gamma_2 d_2} - {\rm e}^{-i{\cal
      E}_2d_2}\r  
\end{array}
\right).
\ee

Applying the evolution operator $U(d)$ one can find the system wave
function after passing as many layers  (let it be $n$) as needed to approximate a 
given experimental setup. The results can be written as
\[
\psi(d)=\l\Pi_{k=n}^1 U_k(d_k)\r\psi(0)
\]
with $d=\sum_{k=1}^{n}d_k$. Consequently, for the probability to find
the hidden photon starting from the visible photon state one finds
\begin{equation}
  \label{multi-layers-probability} 
P(d)=|\psi_2(d)|^2=\left|\l\Pi_{k=n}^{1} U_k(d_k)\r_{12}\right|^2\;.
\end{equation}

\section{Detection of hidden photons: light shining through the wall}
\label{sec:registration}

For detection of the hidden photon the situation is opposite: as initial state
we have a pure hidden photon (gauge eigenstate) and intend to observe a
visible photon. The system starts to evolve in a dense media, where all the
photons are absorbed, and only the hidden photons, singlets with respect
to the SM gauge group and hence immune to interaction with matter,
remain. 

In general case the conversion amplitude is obtained, as usual, by 
solving the corresponding Schrodinger equation with space-dependent
entries in the Hamiltonian. 
The analytic formula for the probability of conversion
$\gamma'\to\gamma$ can be obtained for the case $m_\gamma,\Gamma=\const$
corresponding to a homogeneous media or when the media can be
approximated by a set of homogeneous layers.

In general, there are two types of detectors capable
of hunting the light exotics. One is an empty volume surrounded by
detectors aiming at observation of particles {\it emerging from
  nothing}. The relevant wave function then reads from Eq.\,\eqref{uU}
as 
\be
  \psi(0) \equiv \left(
  \begin{array}{c}
    0 \\
    1
  \end{array}\right) \to
  \psi(d) \approx \left(
  \begin{array}{c}
    - \frac{\delta}{{\cal E}}\left(1 - {\rm e}^{-i{\cal E}d}\right) 
    \\
    {\rm e}^{-i{\cal E}d}
  \end{array}
  \right)
  \equiv U(d)\psi(0)\;,
  \label{vac-inv}
  \ee
which gives \eqref{vacuum-probability} for the conversion probability
$\gamma'\to\gamma$, the same result as for the inverse process.

Another type is dense media with a veto system preventing photon
entering from outside. In this case the wave function reads from
Eq.\,\eqref{uU} as 
\be
  \psi(0) \equiv \left(
  \begin{array}{c}
    0 \\
    1
  \end{array}\right) \to
  \psi(d) \approx \left(
  \begin{array}{c}
    - \frac{\delta}{{\cal E}+i\gamma}\left({\rm e}^{-\gamma d} - {\rm e}^{-i{\cal E}d}\right) 
    \\
    {\rm e}^{-i{\cal E}d}
  \end{array}
  \right)
  \equiv U(d)\psi(0)\;.
  \label{mat-inv}
  \ee
In a realistic setup the propagation distance $d$ is not fixed,
in the sense that the visible photon can interact (and thereby can be
detected) at any point inside the detector of length $L$. Thus to
calculate the probability to detect the photon inside the detector one can
instead calculate the probability of the state to leave the detector
volume. This probability is equal to
\be
P\l\gamma'\to\gamma\r = 1-\psi^\dagger(L)\psi(L)
\ee
and it can be calculated using general formula~\eqref{exact} for
evolution of the wave function. For
$\left|\frac{\delta}{{\cal E}+i\gamma}\right|\ll 1$ one obtains
\be
\label{res-gen}
P\l\gamma'\to\gamma\r \approx
\frac{2\delta^2}{{\cal E}^2+\gamma^2}
\left[\gamma L - \frac{1}{2}\left(1+{\rm e}^{-2\gamma L} -2{\rm e}^{-\gamma L}\cos{{\cal E}L}\right) +
  \frac{{\cal E}^2-\gamma^2}{{\cal E}^2+\gamma^2}\left(1-{\rm
    e}^{-\gamma L}\cos{{\cal E}L}\right) -\frac{2{\cal E}\gamma\,{\rm
      e}^{-\gamma L}\sin{{\cal E}L}}{{\cal
      E}^2+\gamma^2}\right] 
\ee
Below we consider several limiting cases. 
For $\gamma L\gg 1$ and $\left|\frac{\delta}{{\cal
    E}+i\gamma}\gamma L\right|\ll 1$ one has 
\be
\psi(L) \approx \left(
\begin{array}{c}
  \frac{\delta}{{\cal E}+i\gamma} \\
  1 + \frac{\delta^2}{({\cal E}+i\gamma)^2}
\end{array}\right){\rm e}^{-i{\cal E}L -
  \frac{\delta^2\gamma}{{\cal E}^2 + \gamma^2}L}
\ee
and the probability to detect visible photon \eqref{res-gen} is
estimated as 
\be
\label{prob-det}
P\l\gamma'\to\gamma\r = 1 - \psi^\dagger(L)\psi(L) \approx
\frac{2\delta^2\gamma}{{\cal E}^2+\gamma^2}L = \frac{\epsilon^2
  m^4}{(\Delta m^2)^2+E^2\Gamma^2}\Gamma L
\ee
assuming 100\% efficiency of the photon detection. One can see that this
probability is enhanced by a large factor $\Gamma L$ as compared to
expression~\eqref{probability-constant}. 

Another interesting limit corresponds to the low photon absorption,
i.e. $\gamma L\ll 1$ but still $\left|\frac{\delta}{{\cal
    E}+i\gamma}\right|\ll 1$. Using simple algebra one obtains from
\eqref{res-gen} 
\be
P\l\gamma'\r \approx \frac{\delta^2}{{\cal E}^2 + \gamma^2}
4\gamma L \l 1 - \frac{\sin{{\cal E}L}}{{\cal E}L}\r = \frac{\epsilon^2 m^4}{({\Delta
      m^2})^2+E^2\Gamma^2} \l 2\Gamma L\r \l 1 - \frac{\sin{\left(\frac{\Delta
        m^2}{2E}L\right)}}{\left(\frac{\Delta m^2}{2E}L\right)} \r\;.
\ee
Let us note that many of the experimental setups discussed further contain
a veto system. Therefore, the resulting detection probability should be
corrected by a probability to pass the veto.
In all the cases the conversion probability $P\l \gamma' \to \gamma\r$
must be convoluted  over energy with expected hidden photon flux and
weighted with photon detection efficiency along the trajectory inside
the fiducial volume of a given detector.

\section{Example experiments}
\label{sec:examples}

In this Section we discuss prospects of various types of
experiments in probing models with the light hidden
photon. As in neutrino oscillation studies the experiments can be of
``appearance'' and ``disappearance'' types. The former implies that
the hidden photons are produced due to their mixing  with the visible photons
and after that, to be detected, they should be converted back the
ordinary photons (i.e. light shining through the wall signature); here
the signal probability scales as $\epsilon^4$. The latter case assumes
detection of a diluted photon flux or some missing energy signature
and the probability of the photon disappearance scales as
$\epsilon^2$.  

In the following discussion we assume that the regime of
the visible-to-hidden photon oscillations is at work. In a generic
experimental setup the visible photons are produced at some point,
propagate some distance (a part of its trajectory can lie in an absorbing
media) converting into the hidden photons and back and are detected somehow. In the case, when the 
photons are produced in vacua and then cover the distance much 
exceeding the oscillation length, the probability to find the oscillating
system in the hidden photon state
after the absorbing media is given by either Eq.~\eqref{vacuum-probability} or~\eqref{prob-av-vacuum} if we
average over the photon energy spectrum. At the same time, if the photon is produced in
media then the probability to obtain the dark photon outside this
absorbing part of the experiment is determined by
Eqs.~\eqref{general-amplitude} and~\eqref{general-probability} which
for the homogeneous media reduces
to~\eqref{probability-constant}. For heavy hidden photons, namely,
when $m_X\gg m_\gamma$ and $m_X^2\gg E\Gamma$, the probability
~\eqref{probability-constant}  reduces to $\epsilon^2$. In other cases the
probability of hidden photon production gets suppressed. When $\left|
m_X^2-m_\gamma^2\right| \ll E\Gamma$ (high absorption case), the probability is given by 
\be
\label{c1}
P\l\gamma\to\gamma'\r \approx \epsilon^2\frac{m_X^4}{E^2\Gamma^2}.
\ee
In the case $m_X\ll m_\gamma$ and $m_\gamma^2\gg E\Gamma$ (low
absorption case) it is 
\be
\label{c2}
P\l\gamma\to\gamma'\r \approx \epsilon^2\frac{m_X^4}{m_\gamma^4}.
\ee
In both cases the probabilities~\eqref{c1} and~\eqref{c2} scale as
$m_X^4$.
One can see that the photon production in a dense media generally
implies a suppression with respect to the vacuum case which in turn
results in a decrease of the experimental sensitivity to this class of
models. In case of a ``disappearance'' experiment
the expressions~\eqref{probability-constant},~\eqref{c1} and~\eqref{c2} 
determine the experimental sensitivity to the parameters of the hidden
photon model.  

At the same time one can envisage several
experimental setups in which the production of hidden photons can be
enhanced. The first corresponds to the case when the condition $m_X\approx
m_\gamma$ (which we call ``resonance'' in what follows) is satisfied
and the photon absorption is low, $m_\gamma^2\gg E\Gamma$, see
Eq.~\eqref{probability-constant}. 
In this setup, the probability is given by Eq.~\eqref{prob-res} with the resonance
amplification of the hidden photon production. However, for a
given type of media the resonance condition can be fulfilled for only
a single value of hidden photon mass. To cover a wider range of masses
one can imagine a setup where the produced visible photons propagate in a
media with gradually changing density of electrons (MSW-like
transitions) which yields $m_\gamma$ gradually changing along the
photon trajectory.

In the case of ``appearance'' experiments, as we have already mentioned,
the produced hidden photons should be converted back to
ordinary photons which are detected in some process. Actually this
conversion takes place all the way down 
from the absorbing part of experiment to the photon detector. To describe this 
process one can use~\eqref{prob-det}, and
the approximate expressions in low and high absorption cases look
as~\eqref{c1} and~\eqref{c2} with an additional factor $\Gamma L$.
For large
masses of the hidden photon the detection probability is $\epsilon^2 \Gamma L$
assuming 100\% efficiency to detect visible photon (which is
generically determined by the corresponding cross section). 

Resulting probability is a product of the production and detection
probabilities. At large masses of the hidden photon (but still in
the ultrarelativistic regime) it behaves as $\epsilon^4 \Gamma L$. The experimental
bounds in $(m_X,\epsilon)$-plane  are in general weaken to larger
values of $m_X$ due to decrease of 
corresponding production and detection cross sections. At very small masses
of the hidden photon the signal probability scales as $\epsilon^4m_X^8$
and at some value of $m_X$ the experimental bound in
$(m_X,\epsilon)$-plane reaches $\epsilon\sim 1$ and the sensitivity to the
model completely disappears.

\subsection{NA64}
NA64 is a  beam dump  experiment at CERN which uses pure electron beam with the energy
100~GeV~\cite{Gninenko:2013rka,Andreas:2013lya}. The beam hits a
hodoscopic electromagnetic calorimeter (ECAL) which
serves as a target and has a sandwich-like structure. Namely, it
consists of alternating layers of lead (Pb) and scintillator (Sc) each
having 1.5mm thickness. Photons are produced dominantly in the Pb layers via
the bremsstrahlung process. The size of produced photon wave packet can
be estimated as $q^{-1}$, where $q$ is the transferred momentum in this
process. The typical interval of $q$ for the production of
photons with energies much larger than $m_e$ 
is  
\be
\frac{m_e^2}{2E} < q < m_e,  
\ee
with the production cross section saturated at the lower bound. The corresponding size of
the photon wave packet varies in $10^{-11}$--$10^{-5}$~cm interval.  
According to the estimate~\eqref{l_coh} the coherence length can be as
  large as 10~cm, which is within the size of the ECAL for hidden photon masses
 $\lesssim 10^5$\,eV. 
The hidden photon produced in the target 
can carry away significant fraction of the beam energy as they penetrate the rest of the detector 
without significant attenuation. Therefore, their experimental signature is an event  with a large missing
energy in the detector.  The number of such events depends on
the probability of visible-to-hidden photon conversion. The full calculation
of this probability for sandwich-like detector is presented in
Appendix\,\ref{App:Sandwich}. Here we use Eq.~\eqref{ans} to make an estimate for the
limiting cases of heavy and light hidden photons. Below we use
$\Gamma_1^{-1}\sim 0.75$~cm for Pb and $\Gamma_2^{-1}\sim 75$~cm for
Sc layers. The corresponding effective photon masses are estimated as
$m_{\gamma, Pb}\approx 61$~eV and $m_{\gamma, Sc}\approx 21$~eV.

In the case of hidden photon masses $\lesssim 100$~eV (the exact
number depends on the interaction length in Sc) the value ${\cal E}\ll
\gamma$ for both Pb and Sc layers.  Moreover, taking into account that
1) the interaction length in Pb is considerably shorter than that in
the scintillator, i.e. $\Gamma_1^{-1}\ll \Gamma_2^{-1}$, and 2) the
oscillation length~\eqref{osc-length} is longer than the thickness
$d=1.5$~mm of each layer, one can obtain
\be
\label{na64_p_light}
P\l \gamma\to\gamma'\r \approx \frac{4\delta^2}{\gamma_1^2} = 4\epsilon^2\frac{m_X^4}{E^2\Gamma_1^2}\;. 
\ee
So, one finds that the experimental bounds on the models with dark photon
from this experiment scale as $m_X^4$ for $m_X\lsim 100$~eV.

In the case $m_X\gsim 1$~keV and considered photon energies one has
${\cal E}\gg \gamma$ for both types of the layers. In this limit one
obtains from~\eqref{ans} 
\be
\label{na64_p_heavy}
P\l\gamma\to\gamma'\r \approx \frac{\delta^2}{{\cal E}^2}\approx
\epsilon^2\;,
\ee
where ${\cal E}\equiv{\cal E}_1\approx {\cal E}_2$. In this regime the
bound on $\epsilon$ is flat with respect to the mass of the hidden
photon. 

In the case of intermediate masses of dark photon one should apply
the general expression for conversion probability. Comparing
Eqs.~\eqref{na64_p_light} and~\eqref{na64_p_heavy} one can infer that
if $\epsilon_{lim, flat}$ is the experimental bound obtained in the regime of
heavy hidden photons, then for $m_{X}\lsim 100$~eV the bound on the
mixing parameter scales as follows
\be
\label{na64_low}
\epsilon_{lim} \simeq \epsilon_{lim,flat}\frac{E\Gamma_1}{2m_X^2}\;.
\ee
We obtain that the experiment ceases to be sensitive to the model for
$m_X\lsim \sqrt{\epsilon_{lim,flat}E\Gamma_1}$.

\subsection{FASER}
The idea of FASER (ForwArd Search ExpeRiment) is to use forward
physics of the LHC to enrich its discovery potential, see
Refs.~\cite{Feng:2017uoz,Ariga:2018uku}. The detector (an electromagnetic
calorimeter supplemented by a tracking system) of a size about
$\sim$10~m is 
suggested to be placed in the empty tunnel right along the collision
line at 480~m from the interaction point of the ATLAS experiment.
The hidden photon is among the several types of
new physics models which can be explored in this experiment.
In Ref.~\cite{Feng:2017uoz} it was found that the FASER  discovery potential
using lepton pair final state includes previously unprobed region with
dark photon mass $m_X\sim 10$~MeV$-1$~GeV and mixing $\epsilon\sim
10^{-7}-10^{-3}$. For the case $m_X<2m_e$ the hidden photon decay into $e^+e^-$ is
forbidden but still using ``shining through the wall'' type of searches one
can extend the discovery potential to even smaller masses of hidden 
photons.

Forward photons are produced  dominantly in $\pi^0$
decays~\cite{Feng:2017uoz} at the interaction point and have energies
in a wide interval from hundred GeV to few TeV. 
They travel in the beam pipe for a distance of about 130~m and reach 
the TAN absorber of neutral particles. Light hidden photons can be produced
in oscillations in the beam pipe, then travel to the FASER detector
where get converted back to visible photons.
Requiring that the oscillation length~\eqref{osc-length} is
less than corresponding coherence length~\eqref{l_coh} which is larger
than several hundred meters one can find that the oscillation picture
for the description of the photon - hidden photon system is valid for
the mass interval 
\be
\label{mass-range}
m_X\sim 30\;{\rm eV}-30\;{\rm MeV},
\ee
and here we assume the photon energy $E\simeq 100$~GeV for an
estimate.  Probabilities of the
visible-to-hidden photon conversion and the inverse process can be found
using formulas from the previous Section, and the single photon 
appearing in the detector after absorber would be the experimental
signature. A photon produced in $\pi^0$-decay is converted to a
hidden photon on its way to the TAN absorber with the probability 
given by either~\eqref{prob-av-vacuum} or~\eqref{vacuum-probability}. The produced hidden photon propagates
about 350~m to the FASER detector. If the latter will consist of
alternating layers of scintillating and absorbing
materials one can apply the analysis of Section~3 to obtain analytic
formulas describing the visible photon production probability.
As a simple estimate within \eqref{mass-range}, assuming lead as the main absorbing component
(cf. Section~5.1) the signal probability for the hidden photon of
masses 
\be
m_X\gsim 1\;{\rm keV}
\ee
is $\epsilon^4\Gamma L$ where $L\sim 10$~m is the length of the FASER 
detector and $\Gamma^{-1}$ is the photon interaction (or absorption)
length in lead. For lighter hidden photons the probability gets
additionally suppressed by a factor $\frac{m_X^4}{E^2_\gamma\Gamma^2}$
which greatly decreases sensitivity of this type of searches.

\subsection{MATHUSLA}
MATSUSHLA project~\cite{Curtin:2018mvb} has been proposed to search for
long lived particles produced in proton-proton collisions at LHC.
It utilizes a large $200\times200\times20$~m$^3$ detector volume filled
with air whose roof is covered with a multilayer tracker to detect
highly energetic particles emerging inside 'from nothing'. In particular, it is
capable of detecting emerging photons\,\cite{Curtin:2017izq}. In our scenario,
the photons with interesting kinematics are produced mainly by
pions~\cite{Curtin:2018mvb} from the proton-proton collisions. These
photons can be converted into the hidden photons on their way through
the ATLAS or CMS detector, propagate few hundred meters and finally
are converted back to the visible photons inside the MATHUSLA detector
volume.

To estimate the hidden photon production probability one should take
into account the inner structure of the LHC detector. Produced in p-p
collisions photons
pass the tracking system and then get absorbed in an electromagnetic
calorimeter. To simplify the following estimates we treat the tracker
part of the photon path of order 1~m as the vacuum part and the ECAL part
of a length about 0.2~m (we take CMS) as that of filled with uniformly
distributed 
matter. Then, the state of the system after passing through the tracker
and ECAL parts can be described by Eq.~\eqref{vac-red} where indices 1
and 2 correspond to the vacuum and matter cases, respectively. Taking
$\gamma_1=0$ and assuming $\gamma_2d_2\gg 1$ one obtains for the
probability of the hidden photon production
\be
P\l\gamma\to\gamma'\r = \left|-\frac{\delta}{{\cal E}_1}\left(1-{\rm
  e}^{-i{\cal E}_1d_1}\right){\rm e}^{-i{\cal E}_2d_2} +
  \frac{\delta}{{\cal E}_2 + i\gamma_2}{\rm e}^{-i{\cal E}_2d_2}\right|^2.
\ee
For ${\cal E}_1d_1\ll 1$ which corresponds to the case when
the oscillation length much exceeds the vacuum part of the photon
path, the probability is reduced to
\be
P\l\gamma\to\gamma'\r = \frac{\delta^2}{{\cal E}_2^2 + \gamma_2^2}. 
\ee
Hence, for $m\gg m_{crit}\sim 1$~keV$\l\frac{E}{100~{\rm
    GeV}}\r^{1/2}$  the probability transforms into
$P\l\gamma\to\gamma'\r \approx \epsilon^2$ while at $m\ll m_{crit}$
one finds
\be
P\l\gamma\to\gamma'\r \approx \epsilon^2\l\frac{m}{m_{crit}}\r^4.
\ee

One can try to instrument the  MATHUSLA detector with a large area ECAL. 
In this case the signature of the $\g'$ event  would be detection of a 
visible photon in the ECAL appearing from  nothing. 
The direction reconstruction as well as timing of the
events are supposed to reduce the possible background. Corresponding
probability of the hidden-to-visible photon oscillations is given by 
Eq.~\eqref{probability-homogeneous} where $\Gamma$ is the inverse photon
interaction length in air and $L$ is the distance between the photon
enter and exit points in the MATHUSLA detector. Therefore, a direction
dependence of the detection probability is expected. Let us note, that
at not very small values of the mixing parameter $\epsilon$ the propagation of
dark photon in the rock between the production point and the detector can
be important and decrease the expected signal. This
can be taken into account by using the suppression exponent discussed after 
Eq.~\eqref{probability-homogeneous}.

\subsection{SHiP}
SHiP project~\cite{Anelli:2015pba,Alekhin:2015byh} at CERN is planned to use 400 GeV
proton beam from SPS with a thick Molybdenum-Tungsten target and a hadron
stopper made of iron placed behind the target. This suggests that the
hidden photons will be produced in the neutral pion decays in matter and to calculate the
production probability of dark photons one should utilize
Eq.~\eqref{probability-constant}.

The produced hidden photons propagate through a magnet system introduced to
deflect muons and come to a dedicated $\nu_\tau$-neutrino
detector. This installation will be made of alternating bricks of lead and
nuclear emulsion foils 
and can potentially detect a
signal photon through $e^+e^-$ production. The probability of visible photon
production can be calculated similarly to the hidden photon production
probability in a sandwich-like structure presented in Appendix~B. 

\subsection{T2K}
In T2K experiment the near detector~\cite{Abe:2011ks} can be potentially used to search for
the hidden photons. In this case J-PARC proton beam hits a target of
several tens cm  made of graphite. The produced pions decay partly in
matter, partly in vacuum. To calculate the hidden photon production
probability one should use combinations of propagation amplitudes in
vacuum and matter as described in Section~3.

The near neutrino detector (ND280)is located at 280~m away from the beam
dump and contains Pi-Zero Detector  
whose primary goal is to measure a background from neutral pions. It is
a plastic scintillator-based detector consisting of alternating layers
of scintillator planes,  water bags, and brass sheets. The hidden photon
detection probability here can be found using formulas presented in
Appendix~B.

\subsection{DUNE}

DUNE project~\cite{Acciarri:2016crz,Abi:2018dnh} is planning to use 
120 GeV proton beam at Fermilab hitting a thin target. Photons can be
produced in $\pi^0$ decays in a decay pipe of about 205 m long.
The hidden photon production probability can be calculated using
formula~\eqref{vacuum-probability} for oscillations in vacuum averaged
over photon energy distribution. The decay pipe ends with a hadron
absorber where production of secondary pions is possible. 

Near neutrino detector will be located at a distance of about 210~m
from the absorber hall. Its design is not fixed at the moment. But
most probably it will have a part with the sandwich-like structure and
the corresponding detection probability can be calculated as it is
described in Appendix~B.

\subsection{NA62}
NA62~\cite{NA62:2017rwk} is a beam dump experiment at CERN which uses
400 GeV proton beam delivered by SPS. The visible photons would be
produced in $\pi^0$ decays either in the beam target or in the vacuum
pipe just behind it. Then the (photon) beam propagates through a
vacuum tunnel and several structured (i.e. detector systems) filled
with media about several hundred meters. The hidden photon production
probability can be calculated using formula~\eqref{vacuum-probability}
for oscillations in vacuum averaged over the photon energy distribution. 

Conversion back to the visible photons to be detected in ECAL can
happen in the vacuum decay tunnel as well as in the ECAL itself or
nearby material. Corresponding probability can be calculated using
formulas presented in Sec.~4 and Appendix~B. 

\section{Conclusion}
\label{sec:summary}

To summarize, we investigated the production, propagation and detection
of very light stable hidden photons, which can oscillate into visible
photons due to kinetic mixing, in accelerator type of experiments.
In any such a setup there is a range of model parameters where the oscillation
description is valid. Our study can be used to extend the sensitivity of the experiments
to the models with hidden photons lighter than 1 MeV. 
The oscillations, very similar to those of
neutrinos, proceed differently in vacuum and media. As a result,
generically, the experiment capable of searching for the hidden
photon signatures --- missing photon (disappearance) or visible photon
emerging from nothing (appearance) --- gradually looses the
sensitivity to the light hidden photon starting from the hidden photon
mass at least of order the plasma frequency in matter.

\begin{figure}[!htb]
  \centerline{
    \includegraphics[width=0.7\textwidth]{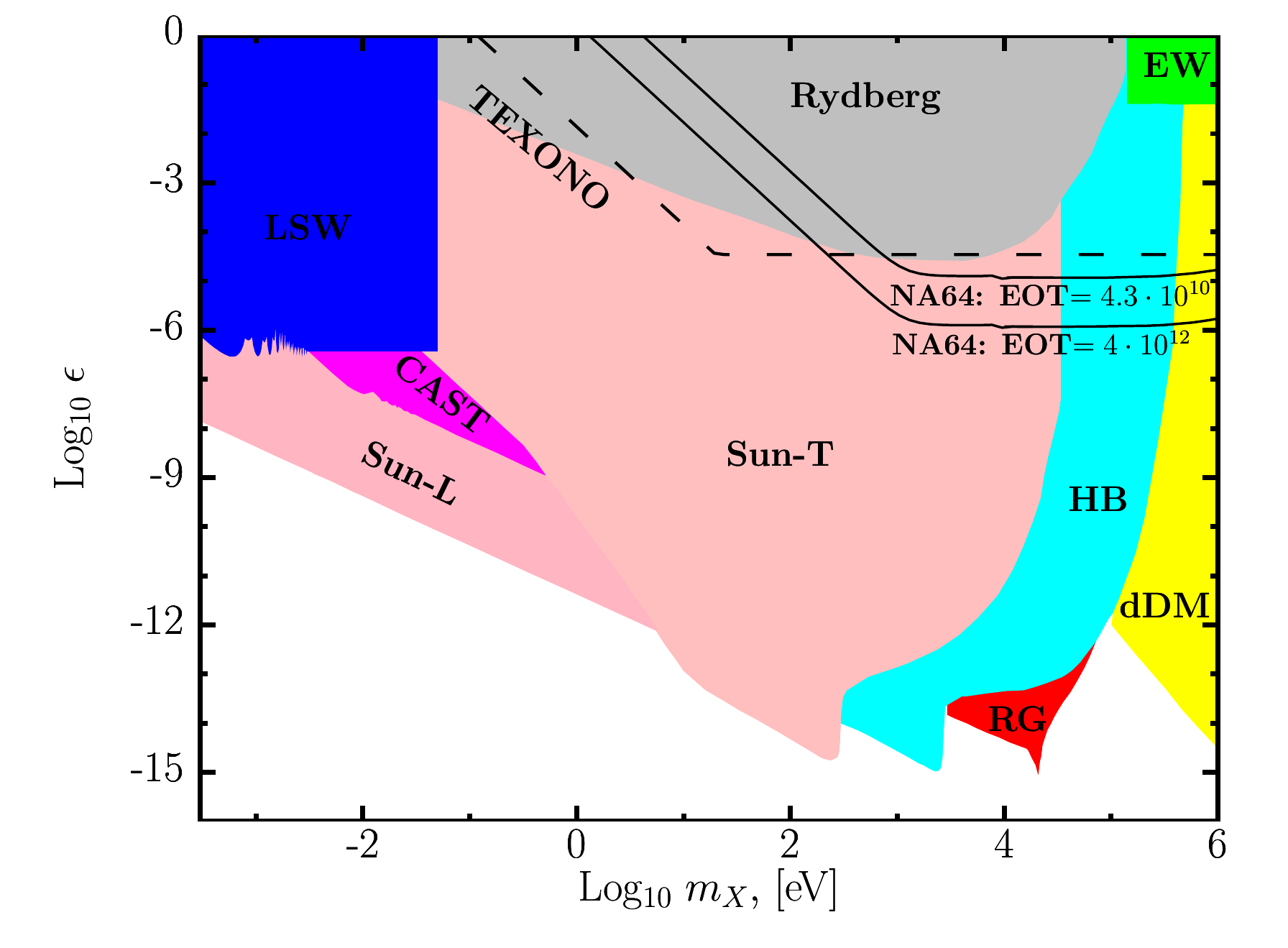}}
  \caption{\label{fig} Exclusion  limits in the $(m_X, \epsilon)$
    parameter plane  of the hidden photon model
    evaluated from the direct searches in the NA64 \cite{Banerjee:2017hhz, Gninenko:2017yus} (solid black lines) and TEXONO \cite{Danilov:2018bks}(dashed black lines) experiments in comparison
    with other  results, see Refs.~\cite{Redondo:2013lna,Hewett:2012ns,An:2014twa} for details. }
\end{figure}

To illustrate our results,  in Fig.~\ref{fig}
we present  limits  expected  from the  NA64 experiment on the model
parameter space in the mass range 1 eV $\lesssim m_X \lesssim$ 1
MeV. The two NA64 limit curves  obtained by making use of
Eq.~\eqref{ans},  refer to the statistics already collected by the
experiment corresponding to $4.3\cdot 10^{10}$ electrons on target
(EOT) Ref.~\cite{Banerjee:2017hhz} and to the ultimate statistics from
the original proposal Ref.~\cite{Andreas:2013lya}. The
mixing-independent horizontal parts match to the sensitivity lines 
presented in Ref~\cite{Gninenko:2017yus} for $m_X>100$~keV,  where the
oscillation formalism  we presented here reduces to the standard
Compton-like description. At small masses the behaviour of the
sensitivity limits matches Eq.~\eqref{na64_p_light}. In Fig.~\ref{fig}
we also outline the exclusion regions evaluated from the results of
the nuclear reactor experiment TEXONO \cite{Danilov:2018bks}, direct
experimental searches (Rydberg, TEXONO, LSW,  EW, and  CAST) and
disfavoured from the astrophysical considerations  (HB, Solar
Lifetime, dDM). For detailed discussion of various limits, see
e.g.~\cite{Danilov:2018bks,Redondo:2013lna,Hewett:2012ns,An:2014twa}.
One may argue that the bounds from stellar cooling are
  considerably more stringent than those from any expected direct
  searches. However, as we discussed in Introduction, results of
  the direct searches are much less sensitive to unknown
  systematics as compared to the astrophysical bounds which are typically not
  assumption-free. As an example,
  we point out that the Standard Solar Model, used to place the bounds
  presented in the Fig.~\ref{fig}, fails to simultaneously explain helioseismic data 
  and photometric observables~\cite{Vinyoles:2016djt}. Unknown
  dynamics (including possible effect of new physics) behind this
  discrepancy could potentially change the solar bounds in
  Fig.~\ref{fig}.
  In addition, using the analysis of the light hidden photon production,
propagation and detection developed in this work the future generation of
the NA64-like experiments, such as e.g. eSPS/LDMX at the CERN SPS~\cite{Akesson:2640784}
which could potentially accumulate more than $10^{17}$ electrons on target, 
will be able to start direct probing the parameter space constrained from
the astrophysical considerations with a much better sensitivity than
NA64.

On the bright sight, there are two advantages inherent in the
oscillation phenomenon. First, the probability to find a single
photon in the detector generally exhibits a dependence on the distance
covered by the oscillating state, see eq.\,\eqref{res-gen}, which can
be exploited to suppress the background and/or distinguish the
hidden-photon model from other SM extensions sharing the same
signature. Moreover, this dependence provides a unique possibility to
pin down the mass of the light hidden photon.

Second, the visible-to-hidden photon transition can be resonantly
enhanced, provided by certain relations between the hidden photon mass
$m_X$, photon effective mass $m_\gamma$, absorption rate $\Gamma$ and
energy $E$:
\begin{equation}
  \label{res-conditions}
  m_X^2=m_\gamma^2\,,\;\;\;\;E\Gamma<m_\gamma^2\,.
\end{equation}
Note, that in a given experiment, even if the energy of primary beam
is too high to fulfill the inequality \eqref{res-conditions}, there
are less energetic particles emergent in the subsequent hadronic
(electromagnetic) cascade, which can produce photons obeying
\eqref{res-conditions}. In particular, for the SHiP experiment,
the relevant photon must come from rather soft, $\sim 320$\,MeV
pions. In the tungsten beam-dump these neutral pions decay into photons which
conversion  into the hidden photons of $m_X\simeq 80$\,eV gets the
resonance amplification.

\vskip 0.3cm
The authors thank V. Duk, Yu. Kudenko and B. Kerbikov for useful
discussions.  
The work is supported by the RSF grant 17-12-01547.

\appendix
\section{Solution of the Schrodinger equation in homogeneous media}
\label{App:Solution}

In a homogeneous media exact solution of the Schrodinger equation with 
the Hamiltonian~\eqref{ham} reads $\psi(L)=\e^{-iHL}\psi(0)$ and can be
explicitly written as 
\be
\label{exact}
\psi(0) \equiv \left(
\begin{array}{c}
  a \\
  b
\end{array}\right) \to
\psi(L) = \left(
\begin{array}{c}
  \left(\frac{1}{2}(1-\alpha)a+\beta b\right){\rm e}^{-i\lambda_+L}+
  \left(\frac{1}{2}(1+\alpha)a-\beta b\right){\rm e}^{-i\lambda_-L}\\
  \left(\beta a+\frac{1}{2}(1+\alpha)\right){\rm e}^{-i\lambda_+L}+
  \left(-\beta a+\frac{1}{2}(1-\alpha)b\right){\rm e}^{-i\lambda_-L}
  \end{array}\right)\;,
\ee
where
\be
\alpha = \frac{{\cal E}+i\gamma}{\sqrt{({\cal
      E}+i\gamma)^2+4\delta^2}}\;\;\;
\beta = \frac{\delta}{\sqrt{({\cal E}+i\gamma)^2+4\delta^2}}\;,
\ee
and $\lambda_{\pm}$ are the Hamiltonian eigenvalues
\be
\lambda_{\pm} = \frac{1}{2}\left({\cal E}-i\gamma\pm\sqrt{({\cal E}+i\gamma)^2+4\delta^2}\right)\;.
\ee
This exact solution can be used for any set of the model and media
parameters. 

\section{Probability to produce the dark photon inside a sandwich-like structure}
\label{App:Sandwich}

We calculate the probability to find the dark photon after a passage through a
sandwich-like structure which is typical for electromagnetic calorimeters.

Let us assume, that the sandwich-like structure consists of two
types 
of layers, of equal widths $d$. One layer is assumed to be filled
with some matter (e.g. lead) which is characterized by $\gamma_1,
{\cal E}_1$, while the other layer characterized by $\gamma_2,
  {\cal E}_2$ is filled with almost vacuum (scintillator), i.e. $\gamma_2\approx 0$ in this case.  

Let us describe the evolution of wave function in each of the layers,
numbered below by a subscript $k=1,2$.
We just make use of the solution \eqref{exact} for each layer and
properly combine them. Below 
we present the results up to corrections of order $\epsilon^2$ (or
$\delta^2$). 
  \be
  \psi(0) \equiv \left(
  \begin{array}{c}
    a \\
    b
  \end{array}\right) \to
  \psi(d) \approx \left(
  \begin{array}{c}
    a{\rm e}^{-\gamma_k d} - \frac{\delta}{{\cal E}_k + i\gamma_k}b\left({\rm
      e}^{-\gamma_k d} - {\rm e}^{-i{\cal E}_kd}\right)
    \\
    b{\rm e}^{-i{\cal E}_kd} - \frac{\delta}{{\cal E}_k +
      i\gamma_k}a\left({\rm e}^{-\gamma_k d} - {\rm e}^{-i{\cal E}_kd}\right)
  \end{array}
  \right)
  \equiv U_k(d)\psi(0)\;,
  \label{u}
  \ee
see also Eq.~\eqref{uU}.
  Here we introduced corresponding evolution matrices $U_1(d)$ and
  $U_2(d)$ for both types of layers.

Now let us find how the wave function of the system changes after
passage through a single pair of layers. The
propagation is described by the matrix $\hat{O}(d)\equiv U_2(d)U_1(d)$,
and we obtain 
\be
\label{1}
\begin{split}
\psi(d)& = U_2(d)U_1(d) \left(
\begin{array}{c}
  a \\
  b
\end{array}\right)
\\&= \left(
\begin{array}{c}
  a{\rm e}^{-\l\gamma_1+\gamma_2\r d}
  \\
  b{\rm e}^{-i\l{\cal E}_1+{\cal E}_2\r d} - \frac{\delta}{{\cal E}_1 +
    i\gamma_1}a\l {\rm e}^{-\gamma_1 d} - {\rm e}^{-i{\cal
      E}_1d}\r {\rm e}^{-i{\cal E}_2d} - \frac{\delta}{{\cal
    E}_2+i\gamma_2}a \l \text{e}^{-\gamma_2 d} - {\rm e}^{-i{\cal
      E}_2d}\r \text{e}^{-\gamma_1 d} 
\end{array}
\right)
\end{split}
\ee
Here we made a simplification: in Eq.~\eqref{1} we omitted a contribution
proportional to $b$ in the ``upper'' element of the wave function as
it is the second order in $\delta$ (remind that we are interested in
the pure photon initial condition).

Now, having in mind the sandwich-like structure of the detector let us
consider Eq.~\eqref{1} as a recurrent relation of the following type
\be
\left(\begin{array}{c}
  a_{n+1} \\
  b_{n+1}
\end{array}
\right)  =
\hat{O}(d) \left(
\begin{array}{c}
  a_n\\
  b_n
\end{array}\right)
=
\left(\begin{array}{cc}
  A & 0 \\
  C & B
\end{array}
\right)
\left(
\begin{array}{c}
  a_n\\
  b_n
\end{array}\right),
\ee
which relates the wave functions at boundaries of a single complex
lead-scintillator layer (which actually consists of two elementary layers).
Here
\be
\label{notations}
\begin{split}
&A = {\rm e}^{-\l\gamma_1+\gamma_2\r d},\;\;\;
B = {\rm e}^{-i\l{\cal E}_1+{\cal E}_2\r d},\;\;\;\\
&C = -\frac{\delta}{{\cal E}_1 + i\gamma_1}\left({\rm e}^{-\gamma_1 d} -
       {\rm e}^{-i{\cal E}_1d}\right){\rm e}^{-i{\cal E}_2d} -
       \frac{\delta}{{\cal E}_2+i\gamma_2}\left(\e^{-\gamma_2 d}-{\rm e}^{-i{\cal E}_2d}\right){\rm
         e}^{-\gamma_1 d}.
\end{split}
       \ee

If we start with pure photon wave function, i.e. $\psi(0) \equiv
\psi_0 = (1\;\; 0)^{\rm T}$, then the asymptotic wave function
is given by the 
following limit
\be
\label{eq_lim}
\psi(\infty) = \lim_{n\to\infty} \hat{O}^n\left(\begin{array}{c}
  1 \\
  0
\end{array}
\right)
\ee
with
\be
\label{eq_lim1}
\hat{O}^n\left(
\begin{array}{c}
  1\\
  0
\end{array}
\right)
= \left(
\begin{array}{c}
  A^n \\
  \frac{C}{A - B}\left(A^n - B^n\right)
\end{array}
\right)
\ee

We see that at $n\to\infty$ the photon part of the wave function goes
to zero, while the probability to find a dark photon is given by
\be
\label{ans}
\begin{split}
P & = \lim_{n\to\infty}\left|\frac{C}{A-B}\left(A^n -
B^n\right)\right|^2 = \left|\frac{C}{A-B}\right|^2 \\&=
\delta^2\left|\frac{\frac{1}{{\cal E}_1 + i\gamma_1}\left(1 -
       {\rm e}^{-i{\cal E}_1d+\gamma_1 d}\right) +
       \frac{1}{{\cal E}_2+i\gamma_2}\left({\rm e}^{i{\cal
           E}_2d-\gamma_2 d}-1\right)}{{\rm e}^{-\gamma_2 d + i {\cal E}_2 d} - {\rm
    e}^{-i{\cal E}_1d+\gamma_1 d}}\right|^2.
\end{split}
\ee
One can easily check that $P\to 0$ if $\Delta m^2\to 0$ as expected. Then
\be
\label{2-layers}
P = \frac{\rm Num}{\rm Den},
\ee
where
\begin{gather}
  {\rm Num} = \frac{\delta^2}{{\cal E}_1^2 + \gamma_1^2}
  \l 1- 2{\rm e}^{\gamma_1d}\cos{{\cal E}_1d}+ {\rm e}^{2\gamma_1
    d}\r 
  \\
+ \frac{\delta^2}{{\cal E}_2^2 + \gamma_2^2}
  \l 1- 2{\rm e}^{-\gamma_2d}\cos{{\cal E}_2d}+ {\rm e}^{-2\gamma_2
    d}\r 
\\
+ \frac{2\delta^2\l {\cal E}_1{\cal E}_2+\gamma_1\gamma_2\r}
{\l{\cal E}_1^2 + \gamma_1^2\r \l{\cal E}_2^2 +
  \gamma_2^2 \r}
\l
-1-{\rm e}^{\l \gamma_1-\gamma_2\r d}\cos{\l {\cal E}_1+{\cal E}_2\r
  d} + {\rm e}^{\gamma_1 d}\cos{{\cal E}_1 d} + {\rm e}^{-\gamma_2 d}\cos{{\cal E}_2 d}
\r
\\
+ \frac{2\delta^2\l \gamma_1{\cal E}_2 - \gamma_2{\cal E}_1\r}
{\l{\cal E}_1^2 + \gamma_1^2\r \l{\cal E}_2^2 +
  \gamma_2^2 \r}
\l
  {\rm e}^{\l \gamma_1-\gamma_2\r d}\sin{\l {\cal E}_1+{\cal E}_2\r d}
  -{\rm e}^{\gamma_1 d}\sin{{\cal E}_1 d}-{\rm e}^{-\gamma_2 d}\sin{{\cal E}_2 d} 
\r
\end{gather}
and
\be
{\rm Den} = ({\rm e}^{\gamma_1 d}-{\rm e}^{-\gamma_2 d})^2 + 4{\rm
  e}^{\l \gamma_1-\gamma_2\r
  d}\sin^2{\frac{{\cal E}_1+{\cal E}_2}{2}d}.
\ee
This calculation assumes that the initial photon is produced at the left edge
of a Pb-layer.

Let us introduce a correction related to the production position. We
will assume that the photon is produced inside some Pb-layer, and in
this first layer its path equals $l$. The above formulas for evolution
through the layers allow us to find the wave function after passage
through the first pair of Pb (of length $l$) and vacuum (of length
$d$) layers.

The propagation of the originally pure photon state in the part
  of the lead layer is
described by the simplified matrix~\eqref{u}
\[
U_1(l)=\begin{pmatrix}
\e^{-\gamma_1 l}  & 0\\
-\frac{\delta}{{\cal E}_1+i\gamma_1}
\l \e^{-\gamma_1 l} - \e^{-i{\cal E}_1 l}\r & \e^{-i {\cal E}_1 l} 
\end{pmatrix},
\]
where $l$ is the penetration depth. For the propagation matrix one
obtains
\[
U_1(l)=U_1(d)\times U_1(l-d)\, 
\]
and if $U_2(d)$ describes the state propagation through the vacuum layer
of depth $d$, then the propagation through a pair of lead and vacuum
layers  is described by $\hat{O}(d)$. Now, we are interested to
calculate 
\[
\hat{O}^{n-1} U_2(d)U_1(l)\psi_0=\hat{O}^n U_1(l-d) \psi_0\,,
\]
where $\psi_0^T=(1,0)$, and $x\equiv d-l$ is the depth of the photon
production inside the lead layer. Then, at large $n$ 
\[
\hat{O}^n=\begin{pmatrix}
A^n  & 0\\
\frac{C}{A-B}\l A^n-B^n\r& B^n 
\end{pmatrix} \rightarrow
\begin{pmatrix}
0  & 0\\
-\frac{C B^n}{A-B} & B^n 
\end{pmatrix} 
\]
and hence
\[
\hat{O}^nU_1(-x)=B^n \begin{pmatrix}
0  & 0\\
-\frac{C \e^{x\gamma_1}}{A-B} - \frac{\delta}{{\cal
    E}_1+i\gamma_1} \l \e^{x\gamma_1}-\e^{i{\cal E}_1x}\r & \e^{i{\cal E}_1x} 
\end{pmatrix} 
\]
Then the probability of transition described by $\hat{O}^nU_1(-x)\psi_0$ is
\[
P(x)=\e^{2x\gamma_1} \left|
\frac{C}{A-B} + \frac{\delta}{{\cal E}_1+i\gamma_1} \l
1-\e^{-x\gamma_1 +i {\cal E}_1x}\r
\right|^2\;.
\]
Let us remind again that here $x$ is the depth of the photon
production inside the same lead production layer. 
At $x=0$ this expression turns into~\eqref{ans}.
This expression for the probability should be  averaged (probably with MC
simulations) over $x$.

Now, this probability can be written as a sum of
eq.\,\eqref{2-layers}$\times\e^{2x\gamma_1}=\e^{2x\gamma_1}\text{Num}/\text{Den}$ and
\[
\delta^2\times\l\frac{\l 1-\e^{-x\gamma_1}\r^2+4 \e^{-x\gamma_1}\sin^2{\frac{{\cal
        E}_1x}{2}}}{{\cal E}_1^2+\gamma_1^2} + 2\times ReN\times ReO
+2\times ImN\times ImO\r \times \e^{2x\gamma_1} ,
\]
where
\begin{align*}
ReN&=\frac{{\cal E}_1\l 1-\e^{-x\gamma_1} \cos{{\cal E}_1x} \r -
  \gamma_1 \e^{-x\gamma_1}\sin{{\cal E}_1x}}{{\cal
    E}_1^2+\gamma_1^2}\\
ImN&=-\frac{\gamma_1\l 1-\e^{-x\gamma_1} \cos{{\cal E}_1x} \r +
  {\cal E}_1 \e^{-x\gamma_1}\sin{{\cal E}_1x}}{{\cal
    E}_1^2+\gamma_1^2}
\end{align*}
and $ReO=(G-F)/J$, $ImO=(K+T)/J$ with 
\begin{align*}
J&=\e^{-2\gamma_2 d}+\e^{2\gamma_1 d}-2 \e^{\l\gamma_1-\gamma_2\r d}
  \cos{\l {\cal E}_1+{\cal E}_2\r d}= \text{Den}
\\
G&=\frac{\e^{\l\gamma_1-\gamma_2\r d}\l \gamma_1 \sin{\l {\cal
    E}_1+{\cal E}_2\r d} - {\cal E}_1  \cos{\l {\cal E}_1+{\cal
    E}_2\r d} \r  +
\e^{-\gamma_2d} \l {\cal E}_1 \cos{{\cal E}_2d} - \gamma_1  \sin{{\cal
    E}_2d}\r
- \e^{\gamma_1d} \l {\cal E}_1 \cos{{\cal E}_1d} + \gamma_1  \sin{{\cal
    E}_1d}\r
+{\cal E}_1 \e^{2\gamma_1d} 
}
{{\cal E}_1^2+\gamma_1^2} \\
F&=\frac{\e^{\l\gamma_1-\gamma_2\r d}\l \gamma_2 \sin{\l {\cal
    E}_1+{\cal E}_2\r d} + {\cal E}_2  \cos{\l {\cal E}_1+{\cal
    E}_2\r d} \r  +
\e^{-\gamma_2d} \l {\cal E}_2 \cos{{\cal E}_2d} - \gamma_2  \sin{{\cal
    E}_2d}\r
- \e^{\gamma_1d} \l {\cal E}_2 \cos{{\cal E}_1d} + \gamma_2  \sin{{\cal
    E}_1d}\r
-{\cal E}_2 \e^{-2\gamma_2d} 
}
{{\cal E}_2^2+\gamma_2^2} \\
K&= \frac{\e^{\l\gamma_1-\gamma_2\r d}\l {\cal E}_1 \sin{\l {\cal
    E}_1+{\cal E}_2\r d} +  \gamma_1 \cos{\l {\cal E}_1+{\cal
    E}_2\r d} \r  -
\e^{-\gamma_2d} \l \gamma_1 \cos{{\cal E}_2d} + {\cal E}_1  \sin{{\cal
    E}_2d}\r
+ \e^{\gamma_1d} \l \gamma_1  \cos{{\cal E}_1d} - {\cal E}_1 \sin{{\cal
    E}_1d}\r
-\gamma_1 \e^{2\gamma_1d} 
}
{{\cal E}_1^2+\gamma_1^2} \\
T&=\frac{\e^{\l\gamma_1-\gamma_2\r d}\l \gamma_2 \cos{\l {\cal
    E}_1+{\cal E}_2\r d} - {\cal E}_2  \sin{\l {\cal E}_1+{\cal
    E}_2\r d} \r  +
\e^{-\gamma_2d} \l {\cal E}_2 \sin{{\cal E}_2d} + \gamma_2  \cos{{\cal
    E}_2d}\r
+ \e^{\gamma_1d} \l {\cal E}_2 \sin{{\cal E}_1d} - \gamma_2  \cos{{\cal
    E}_1d}\r
-\gamma_2 \e^{-2\gamma_2d} 
}
{{\cal E}_2^2+\gamma_2^2} 
\end{align*}

So the probability reads
\[
\e^{2x\gamma_1}\times\frac{\text{Num}}{\text{Den}}+
\frac{\delta^2}{\l{\cal E}_1^2+\gamma_1^2\r\text{Den}} 
\l\l
\l 1-\e^{x\gamma_1}\r^2+4 \e^{x\gamma_1}\sin^2{\frac{{\cal
        E}_1x}{2}}\r\times\text{Den}+2\l
\e^{2x\gamma_1}-\e^{x\gamma_1} \cos{{\cal E}_1x} \r \times V
-2\e^{x\gamma_1}\sin{{\cal E}_1x}\times W 
\r
\]
where
\begin{align*}
V&=-\e^{\l\gamma_1-\gamma_2\r d}\cos{\l {\cal E}_1+{\cal E}_2\r d}
+ \e^{-\gamma_2 d}\cos{{\cal E}_2 d}
- \e^{\gamma_1 d}\cos{{\cal E}_1 d}+\e^{2\gamma_1 d}
\\&+\frac{{\cal E}_1{\cal E}_2+\gamma_1\gamma_2}{{\cal E}_2^2+\gamma_2^2}
\l
\e^{-2\gamma_2d}+\e^{\gamma_1d}\cos{{\cal E}_1
  d}-\e^{-\gamma_2d}\cos{{\cal E}_2 d} - \e^{\l\gamma_1-\gamma_2\r d}\cos{\l {\cal E}_1+{\cal E}_2\r d}
\r\\
&+\frac{\gamma_1{\cal E}_2-\gamma_2{\cal E}_1}{{\cal E}_2^2+\gamma_2^2}
\l
\e^{\l\gamma_1-\gamma_2\r d}\sin{\l {\cal E}_1+{\cal E}_2\r
  d}-\e^{-\gamma_2d}\sin{{\cal E}_2 d} -\e^{\gamma_1d}\sin{{\cal E}_1 d}  
\r
\end{align*}
and
\begin{align*}
W&=\e^{\l\gamma_1-\gamma_2\r d}\sin{\l {\cal E}_1+{\cal E}_2\r d}
- \e^{-\gamma_2 d}\sin{{\cal E}_2 d}
- \e^{\gamma_1 d}\sin{{\cal E}_1 d}
\\&+\frac{{\cal E}_1{\cal E}_2+\gamma_1\gamma_2}{{\cal E}_2^2+\gamma_2^2}
\l
\e^{\gamma_1d}\sin{{\cal E}_1
  d}+\e^{-\gamma_2d}\sin{{\cal E}_2 d} - \e^{\l\gamma_1-\gamma_2\r d}\sin{\l {\cal E}_1+{\cal E}_2\r d}
\r\\
&+\frac{\gamma_1{\cal E}_2-\gamma_2{\cal E}_1}{{\cal E}_2^2+\gamma_2^2}
\l
-\e^{\l\gamma_1-\gamma_2\r d}\cos{\l {\cal E}_1+{\cal E}_2\r
  d}-\e^{-\gamma_2d}\cos{{\cal E}_2 d} +\e^{\gamma_1d}\cos{{\cal E}_1 d}+ \e^{-2\gamma_2 d} 
\r
\end{align*}

Finally, we perform averaging over the scintillator layer by making
use of the formulas
\begin{align*}
d^{-1}\int_0^d dx\,\e^{2x\gamma_1} & = \frac{\e^{2\gamma_1d}-1}{2\gamma_1
  d} \equiv I_1 \,,\\
d^{-1}\int_0^d dx\l\l \e^{x\gamma_1}-1\r^2+4 \e^{x\gamma_1}\sin^2{\frac{{\cal
      E}_1x}{2}}\r & = 1+\frac{\e^{2\gamma_1d}-1}{2\gamma_1 d}+
\frac{2\gamma_1 - 2\e^{\gamma_1d}\l \gamma_1 \cos{{\cal
      E}_1d}+{\cal E}_1 \sin{{\cal E}_1d}\r }{\l{\cal E}_1^2+\gamma_1^2 \r d}
\equiv 1+ I_1+2 J_1\,,
\\
d^{-1}\int_0^d dx\l\e^{2x\gamma_1}- \e^{x\gamma_1}\cos{{\cal
    E}_1x}\r & =\frac{\e^{2\gamma_1d}-1}{2\gamma_1 d}+
\frac{\gamma_1 - \e^{\gamma_1d}\l \gamma_1 \cos{{\cal
      E}_1d}+{\cal E}_1 \sin{{\cal E}_1d}\r }{\l{\cal
    E}_1^2+\gamma_1^2 \r d}
=I_1+J_1\,,
\\
d^{-1}\int_0^d dx\,\e^{x\gamma_1}\sin{{\cal
    E}_1x} & = \frac{{\cal E}_1 + \e^{\gamma_1d}\l \gamma_1 \sin{{\cal
      E}_1d}-{\cal E}_1 \cos{{\cal E}_1d}\r }{\l{\cal E}_1^2+\gamma_1^2 \r d}
\equiv J_2\,.
\end{align*}
and find for the averaged probability
\[
P_{av}=I_1\times\frac{\text{Num}}{\text{Den}} + \frac{\delta^2}{\l{\cal
    E}_1^2+\gamma_1^2\r\text{Den}}  \l \l 1+ I_1+2J_1\r
\times\text{Den} + 2V \l I_1+J_1\r - 2 W J_2\r\,.
\]

\bibliographystyle{JHEP}
\bibliography{dphoton-G}

\end{document}